\documentclass[12pt,preprint]{aastex}
\usepackage{multirow}
 \newcommand{\beq}{\begin{equation}}
 \newcommand{\eeq}{\end{equation}}
 \newcommand{\beqn}{\begin{eqnarray}}
 \newcommand{\eeqn}{\end{eqnarray}}
 \newcommand{\non}{\nonumber}
 \newcommand{\no}{\noindent}

\newcommand{\pa}{\partial}

\begin{document}
\title{Galaxies with Supermassive Binary Black Holes:\\
(II) A Model with Cuspy Galactic Density Profiles}
          
\author{Ing-Guey Jiang$^{1}$ and Li-Chin Yeh$^{2}$}

\affil{
{$^{1}$ Department of Physics and Institute of Astronomy,}\\
{ National Tsing-Hua University, Hsin-Chu, Taiwan}\\ 
{$^{2}$ Department of Applied Mathematics,}\\
{ National Hsinchu University of Education, Hsin-Chu, Taiwan} 
}

\authoremail{jiang@phys.nthu.edu.tw}

\begin{abstract} 

The existence and uniqueness of equilibrium points,
including Lagrange Points and Jiang-Yeh Points, 
of a galactic system with supermassive binary black holes embedded
in a centrally cuspy galactic halo are investigated herein.
Differing from the previous results of non-cuspy galactic 
profiles that Jiang-Yeh Points
only exist under a particular condition,
it is found here that the Lagrange Points, L2, L3, L4 and L5,
Jiang-Yeh Points, JY1 and JY2, exist under general conditions.
The stability analysis shows that L2, L3, JY1 and JY2 are unstable.
However, L4 and L5 are only unstable
when the galactic total mass is smaller than a critical mass;
otherwise they become neutrally stable centers.
These results will be important for  further studies on the
cores of early-type galaxies.
 
\end{abstract}

\newpage
\section{Introduction}

The general structure 
of galaxies has been one of the most 
interesting astronomical subjects due to the
beauty and diversity of their morphological shapes.
It is also an important research field in 
that galaxies are building blocks of the universe;
so the formation and evolution of galactic structures, which
are imprints left by the interactions of galaxies, 
can trace the history of the universe.

Through astronomical observations,
the interaction between galaxies is confirmed 
to be taking place frequently, for example, 
the mergers of mice galaxies (NGC 4674 A\&B),
NGC 2207/IC 2163 and NGC 2623.
These results have triggered 
many further investigations related to mergers.
For example, Wu \& Jiang (2009) showed that
the inter-galactic population could be
ejected from galaxies during merging events. 
Furthermore, the ring galaxies are likely to be the outcome
of major mergers, as shown in Wu \& Jiang (2012). 

On the other hand, it is generally believed that there is a supermassive
black hole at the center of galaxies.
A merger of two galaxies with supermassive black holes 
is likely to form a supermassive binary black hole
(SBBH) near the center of the newly formed galactic merging system.  
In fact, the major mergers of two disk galaxies are believed
to be the main mechanism forming elliptical galaxies.
This is partially due to the fact that elliptical galaxies are 
generally massive and
gas-poor, and major mergers can produce a larger
galactic system and reduce the gaseous component by rapid star
formation or gas diffusion during merging processes. 
Thus, the investigations on the dynamics of elliptical galaxies
are usually based on the gas-poor assumption.

In order to study the dynamic evolution of SBBH,
Quinlan (1996) carried out numerical experiments on the scattering
processes for the restricted three-body problem. 
The main issue addressed in that work was the SBBH hardening.  
In fact, modifications of the restricted three body 
problem were employed to model various problems related to  
planetary systems 
(Chermnykh 1987; Papadakis 2004, 2005a, 2005b, 
Jiang \& Yeh 2006; Yeh \& Jiang 2006;
Kushvah 2008a, 2008b, 2009, 2011a, 2011b, 2012).
In addition, the discoveries of extra-solar planets 
have triggered many investigations
in the field of planetary systems (see Jiang \& Ip 2001; Ji et al. 2002;
Jiang \& Yeh 2003, 2004a, 2004b, 2004c, 2007, 2009, 2011;
Jiang et al. 2003, 2006, 2007, 2009, 2010, 2013;
Chatterjee et al. 2008).

Milosavljevic and Merritt (2001)
used N-body simulations to investigate
the orbital decay of black holes and 
the formation of SBBHs during galactic mergers.
They found that it only takes about a million years 
for black holes to sink to the center and become a hard binary.
However, they claimed that black holes stall at separations of
sub-parsec scales.  
Then, Yu (2002) semi-analytically estimated the possible time-scales
to form SBBHs and found that the 
time-scale of dynamical friction is too long for small black holes to 
sink into the center, so it is difficult to form 
SBBHs with very small mass ratio.  
Thus, the SBBHs with moderate mass ratios are most likely to form and survive 
with semi-major axes around $10^{-3}$ to $10$ pc 
in spherical or nearly spherical galaxies. 

Moreover, in order to explain the surface brightness of NGC 3706,
Kandrup et al. (2003) considered the stellar dynamics 
under the gravitational fields of SBBH and a fixed galactic potential.
They found that the transition between the inner and outer power-law
profiles predicted by a Dehnen potential is too gradual
to represent real galaxies, so 
the Nuker law was used in their model.
  
Motivated by the above work, as in Kandrup et al. (2003),
a model of modified restricted three body problems was
used in Jiang \& Yeh (2014) to study a galactic system with SBBH. 
Both Kandrup et al. (2003) and Jiang \& Yeh (2014)
employed the Nuker law as their three dimensional spherical 
galactic density profiles because the Nuker law 
is a simple and neat way to present a smooth broken power-law with 
two power-indexes, and the transition is located at a well defined
radius, i.e. the break radius.  
Note that the Nuker law was originally introduced  
in Lauer et al. (1992) to fit the two dimensional surface brightness
of M32, and later for other galaxies in Lauer et al. (1995).
In order to make it clear, we will use the term {\it spherical Nuker law} 
to name the galactic density profiles in this paper.

Jiang \& Yeh (2014) discovered that when the galactic density profile is 
a spherical Nuker law with $\gamma=0$ (i.e. no cusp),  
$\alpha=2$, $\beta=4$, the usual five Lagrange Points always exist,
and the new equilibrium points, i.e. Jiang-Yeh Points, exist if 
and only if the total galactic mass is larger than the critical mass. 
The analytic expression of this critical mass was provided 
and the stability analysis was performed. These results give important
implications for the orbital evolution near the centers 
of galaxies with SBBH, and lead to a possible mechanism to form
the cores of early-type galaxies through the existence of unstable 
Jiang-Yeh Points near supermassive black holes.  
 
However, because the cuspy density profile is expected 
for the dark matter halo through  
cosmological simulations, it will be interesting to investigate the 
existence of Jiang-Yeh Points in a system with SBBH and 
a central cusp. 
For spherical Nuker law, i.e. Eq.(6) of Jiang \& Yeh (2014), 
the density profile has a cuspy center when $\gamma \ge 1$. 
We find that when $\alpha=2$, $\beta=5$, $\gamma=1$, the corresponding 
gravitational potential of the density profile has an analytic form.
Therefore, in this paper, we study the dynamics of a system wherein a test
particle moves under the influence of SBBH and a galactic potential, 
which is dominated by a cuspy dark matter halo following 
a spherical Nuker law with $\alpha=2$, $\beta=5$, $\gamma=1$.
We will investigate the existence of equilibrium points and 
also perform the stability analysis of these points. 

We present our model in Section 2, and
the analytical results for the existence of equilibrium points 
are in Section 3.
The bifurcation diagrams, locations of equilibrium points  
and zero-velocity curves are shown in Section 4. 
The stability analysis is presented in Section 5, and
the concluding remarks are offered in Section 6.
  
\section{The Model}

We consider the motion of test particles influenced by 
the gravitational force from the central binary black holes and the 
galaxy.
The test particles are considered to move on the same plane of the
orbital plane of binary black holes.

For the description below, those parts which are exactly the same
as the equations in Jiang \& Yeh (2014) will be skipped (refer to Jiang \& 
Yeh 2014).
After the procedure of non-dimensionalization as done 
in Yeh et al. (2012) and Jiang \& Yeh (2014), in which the scale length 
is set to be the break radius, i.e. $L_0 = r_b$,
the density profile of the galaxy is expressed as:
\beq
\rho=\rho_c  r^{-1}\left\{1+ r^{2}\right\}^{-2}\,.\label{rho}
\eeq
 where 
$\rho_c$ is a constant; $r$ is the distance (in the unit of 
break radius $r_b$) from the origin in 
this spherical distribution. 
This profile gives an NFW central cusp (Navarro, Frenk, White 1995)
and also a realistic sharper outer edge, 
i.e. when $r \to 0$, the density $\rho \propto r^{-1}$ is 
the cusp in NFW profile;
when $r >> 1$, the density decays as $\rho \propto r^{-5}$ 
which is sharper than
the NFW's profile ($\rho \propto r^{-3}$ for $r >>1$). 
Because NFW profile is not for an equilibrium 
galaxy (Binney \& Trmaine 2008), our profile 
is a more realistic model for the system considered here. 
Note that all variables, functions and parameters, 
such as mass, potential, radius, normalization constants etc., are all 
written as dimensionless quantities in this paper.

The mass of the galaxy up to $r$ is thus:
\beq
M(r) = 2\pi\rho_c\left\{1-\frac{1}{1+r^2}\right\}.
\eeq
If the total galactic mass is $M_g$, we have $\rho_c=\frac{M_g}{2\pi}$. 
The corresponding potential is:
\beq
V(r)=-4\pi\left[\frac{1}{r}\int_0^r \rho(r') {r'}^2dr' 
+\int_r^\infty \rho(r')r' dr'\right]
=-M_g\left\{\frac{\pi}{2}-\tan^{-1}r\right\},\label{eq:va1}
\eeq
Moreover, from the potential, we can obtain the gravitational force as:
\beq
f_g(r)\equiv -\frac{\pa V}{\pa r}= -M_g\left(\frac{1}{1+r^2}\right).
\label{eq:fg}
\eeq

Considering the motion on the $x-y$ plane 
of the rotating frame, binary black holes, each with mass $m$ 
at $(R, 0)$, $(-R, 0)$, and a test particle
located on $x,y$ with velocity $u$, $v$, 
the equations of motion would be as follows: 
\beq \left\{
\begin{array}{ll}
&  \frac{dx}{dt}=u  \\
& \frac{dy}{dt}=v  \\
& \frac{du}{dt}=2nv +n^2 x-\frac{m(x+R)}{r_1^3}-\frac{m(x-R)}
{r_2^3}-\frac{M_g x}{r}\left(\frac{1}{1+r^2}\right),  \\
& \frac{d v}{dt}=-2nu+n^2 y-\frac{my}{r_1^3}-
\frac{my}{r_2^3}-\frac{M_g y}{r}\left(\frac{1}{1+r^2}\right),
\end{array}  \right. \label{eq:3body3_non}
\eeq
where $r_1 = \sqrt{(x+R)^2+y^2 }$, 
$r_2 =\sqrt{(x-R)^2+y^2 }$, 
$r=\sqrt{x^2 + y^2}$ and $n$ is the angular velocity of black holes.
That is, in the inertial frame, each black hole 
moves along the circular orbit at $r=R$ with an 
angular velocity, i.e. mean motion:
\beq
n=\left\{\frac{m}{4R^3}+\frac{1}{R}|f_g(R)|\right\}^{1/2}=
\left\{\frac{m}{4R^3}+\frac{M_g}{R(1+R^2)}\right\}^{1/2}
\label{eq:n2}
\eeq
The corresponding Jacobi integral of this system is similar as
the one in Jiang \& Yeh (2006, 2014):
\beq
C_J =  -u^2 -v^2 + n^2 (x^2+y^2) 
       + \frac {2m}{r_1} + \frac {2m}{r_2}
       + 2M_g \left\{\frac{\pi}{2}-\tan^{-1}(\sqrt{x^2+y^2})\right\}. 
\label{eq:Jacobi}
\eeq

\section{The Equilibrium Points}

The existence and uniqueness of equilibrium points 
are investigated in this section. 
We follow the convention introduced in Jiang \& Yeh (2014) 
in naming the equilibrium points except that 
due to the singular property of the central point of the system,
the origin will be named as {\it the singular point}. 
Because the singular point is not considered as an
equilibrium point in our system, there will be no stability analysis of
this point.  
The existence and uniqueness of Lagrange Points L4 and L5 will 
be presented in Theorem 1. The existence and uniqueness 
of Lagrange Points L2 and L3, 
and the singularity of the origin of the system will be presented in
Theorem 2. The existence and uniqueness of Jiang-Yeh Points JY1 and JY2,
will be presented in Theorem 3.  
    
In general, for System 
(\ref{eq:3body3_non}), equilibrium points $(x_e,y_e)$ 
satisfy $A(x_e,y_e)=0$ and $B(x_e,y_e)=0$, where: 
\beqn
&&A(x,y)= n^2 x-\frac{m(x+R)}{r_1^3}-\frac{m(x-R)}{r_2^3}-\frac{M_g x}{r}
\frac{1}{(1+r^2)},  
\label{eq:gf2}\\
& & B(x,y)=n^2 y-\frac{m y}{r_1^3}-\frac{m y}{r_2^3}
-\frac{M_g y}{r}\frac{1}{(1+r^2)}. \label{eq:gg2}
\eeqn
For convenience, for $y \ne 0$, we define:
\beq
h(y) \equiv  \frac{B(0,y)}{y}= n^2-\frac{2m}{\left[R^2+y^2\right]^{3/2}}
-\frac{M_g}{|y|}\frac{1}{(1+y^2)}
\label{eq:hye}
\eeq 
and
\beq
k(x) \equiv A(x,0)= n^2x-\frac{m(x+R)}{|x+R|^3}-\frac{m(x-R)}{|x-R|^3}-\frac{M_g x}{|x|}\frac{1}
{(1+x^2)}
 \label{eq:kxe}
\eeq
From the results in Jiang \& Yeh (2006), we have the following remarks:\\
\no{ {\bf Remark A:} }\\
\no {
For $y_e \ne 0$,  $y_e$ satisfies  $h(y)=0$, if and only if 
$(0,y_e)$  is 
the equilibrium point of System (\ref{eq:3body3_non}).\\
{\bf Remark B:} \\
$x_e$ satisfies $k(x)=0$, if and only if
$(x_e,0)$  is  the equilibrium point 
of System (\ref{eq:3body3_non}).
}\\
Then, Remark (A) will be used to study the equilibrium points
L4 and L5 in Theorem 1. Remark (B) will be used 
for all other points in Theorem 2 and 3. 

{\bf Theorem 1: The Existence and Uniqueness of Lagrange Points L4 and L5}\\
{\it There is one and only one 
$\bar{y_1}>0$ such that $h(\bar{y_1})=0$, and only one 
$\bar{y_2}<0$ such that $h(\bar{y_2})=0$. 
That is, excluding the origin (0,0) of $x-y$ plane,
there are two equilibrium points on the $y$-axis, i.e. L4 and L5,
for System (\ref{eq:3body3_non}).}

\no{ {\it Proof:} } 

We define
\beq
P(y)\equiv \frac{2m}{\left[R^2+y^2\right]^{3/2}}-n^2 \label{eq:p1}
\eeq
 \beq
{\rm and} \qquad Q(y)\equiv \frac{-M_g}{|y|(1+y^2)}
\label{eq:q1}
\eeq 
so from Eq.(\ref{eq:hye}), we have $h(y)= -P(y)+Q(y)$.

At first, we consider the case when $y>0$. 
Because $\lim_{y\to 0}Q(y)=-\infty$, we have:
\beq
\lim_{y\to 0}h(y)=\lim_{y\to 0}-P(y)+Q(y)=-\infty;
\eeq
and since $\lim_{y\to \infty}P(y)=-n^2$, $\lim_{y\to \infty}Q(y)=0$,
we have  $\lim_{y\to\infty} h(y)=n^2>0$. 
Moreover, from Eqs.(\ref{eq:p1})-(\ref{eq:q1}),  we have:
\beq
P'(y)= -6my\left[R^2+y^2\right]^{-5/2}, \label{eq:dp1}
\eeq
 and 
\beq
Q'(y)= M_g\frac{1+3y^2}
{y^2(1+y^2)^2}.
\label{eq:dq1}
\eeq
Since $P'(y)<0$ and $Q'(y)>0$, $h'(y)=-P'(y)+Q'(y)>0$ for 
any $y>0$. Thus, $h(y)$ is a monotonically increasing  
function for any $y\in (0, \infty)$.
With $\lim_{y\to 0}h(y)=-\infty$ and $\lim_{y\to\infty} h(y)>0$,
we conclude 
there is a unique point $\bar{y_1}>0$ such  
that $h(\bar{y_1})=0$.

For the case when $y<0$, from Eqs.(\ref{eq:dp1})-(\ref{eq:dq1}), we have 
$P'(y)>0$,$Q'(y)<0$, so $h(y)$ is a monotonically decreasing function. 
We also have $\lim_{y\to\-\infty}h(y)=n^2>0$ and $\lim_{y\to 0}h(y)=-\infty$.
Thus, we find that there is  a unique point $\bar{y_2}<0$ such that 
$h(\bar{y_2})=0$. $\Box$

Now we investigate the existence of  equilibrium points on the $x$-axis. 
For convenience, we define  
\beq
S(x)\equiv \frac{m(x+R)}{|x+R|^3}+\frac{m(x-R)}
{|x-R|^3}-n^2x, \label{eq:p2}
\eeq
and therefore,
\beq
S(x) =\left\{
\begin{array} {lll}
&  \frac{m}{(x+R)^2}+
\frac{m}{(x-R)^2}-n^2x,  &\quad {\rm for} \,\, x>R, \\
& \frac{m}{(x+R)^2}-\frac{m}{(x-R)^2}-n^2x, & \quad {\rm for} \,\, -R<x<R,\\
& -\frac{m}{(x+R)^2}-
\frac{m}{(x-R)^2}-n^2x, & \quad {\rm for} \,\, x<-R.
\end{array}\right. \label{eq:p2_3}
\eeq
We also define:
\beq
T(x)\equiv -\frac{M_g x}{|x|(1+x^2)}.\label{eq:tx}
\eeq
From Eqs.(\ref{eq:kxe}), (\ref{eq:p2}) 
and (\ref{eq:tx}), we have:
\beq
k(x)= -S(x)+T(x). \label{eq:ak}
\eeq

In the following Theorem 2, the results 
show that Lagrange Points L2 and L3 exist. 
However, due to the singular property, 
the origin (0,0) is a singular point, though the net force 
shall be physically balanced there.
In Theorem 3, it is shown that
another two equilibrium points, JY1 and JY2, exist. 
JY1 is in the region
$(-R,0)$ and JY2 is in the region $(0,R)$ of x-axis.

{\bf Theorem 2: The Existence and Uniqueness of Lagrange Points L2 and L3
and the Singularity of the Origin Point (0,0)}\\ 
(i) There is an unique $x_1>R$ such that $k(x_1)=0$
and an unique $x_2<-R$ such that $k(x_2)=0$. 
That is, on the $x$-axis, there is one and only one equilibrium point
in the region $(R,\infty)$, i.e. L2, and there is one and only one
equilibrium point in the region $(-\infty, -R)$, i.e. L3. 
\\
(ii) Due to the singularity at $x=0$, the origin is not 
     an equilibrium point (i.e. $k(0)\ne 0$), but is a singular point.\\

\no {\it Proof of (i):} 

When $x>R$, from Eqs. (\ref{eq:p2_3})-(\ref{eq:ak}), 
$$
k(x)=-S(x)+T(x) 
= -\frac{m}{(x+R)^2}-\frac{m}{(x-R)^2}+n^2x-\frac{M_g}{1+x^2}
$$
$$
{\rm and}\quad 
k'(x)=-S'(x)+T'(x) 
= \frac{2m}{(x+R)^3}+\frac{2m}{(x-R)^3}+n^2+\frac{2xM_g}{(1+x^2)^2}>0.
$$
Thus, we have 
$\lim_{x\to R^{+}} k(x)=-\infty$,
$\lim_{x\to \infty} k(x)=\infty$.
Moreover, from $k'(x)>0$,
we know that $k(x)$ is a monotonic function.
Therefore, there is a unique $x_1>R$, such that $k(x_1)=0$.

When $x<-R$, similarly, since $k'(x)>0$, 
$\lim_{x\to -R^{-}}k(x)=\infty$, and $\lim_{x\to -\infty}k(x)=-\infty$,
there is a unique $x_2<-R$ 
such that $k(x_2)=0$. $\Box$

\no {\it Proof of (ii):}

When $-R<x<R$, from Eq. (\ref{eq:p2_3}), we have:
\beq
S(x) =\frac{m}{(x+R)^2}-\frac{m}{(x-R)^2}-n^2x,
\label{eq:p2_n0}
\eeq
Thus, $S(0)=0$. Moreover, from Eq.(\ref{eq:tx}), we have :
$$
\lim_{x\to 0^{+}}T(x)
=\lim_{x\to 0^{+}}-\frac{M_g}{(1+x^2)}=-M_g$$
and 
$$\lim_{x\to 0^{-}}T(x)
=\lim_{x\to 0^{-}}-\frac{M_gx}{(-x)(1+x^2)}=M_g,$$ 
so $\lim_{x\to 0} T(x)$ does not exist. Therefore, $k(x)$ is not a continuous 
function at $x=0$, and so the origin (0,0) is a singular point.  
$\Box$

{\bf Theorem 3: The Existence and Uniqueness of Jiang-Yeh Points JY1 and JY2}\\ 
There is a unique $x_3\in (0,R)$ such that $k(x_3)=0$ 
and a unique $x_4\in (-R,0)$ such that $k(x_4)=0$.
That is, on the $x$-axis, there is one and only one equilibrium point
in the region $(0,R)$, i.e. JY1, and there is one and only one
equilibrium point in the region $(-R,0)$, i.e. JY2.\\

\no {\it Proof:}

When $-R<x<R$, from Eq.(\ref{eq:p2_n0}), we have:
\beq
S'(x) =-\frac{2m}{(x+R)^3}+\frac{2m}{(x-R)^3}-n^2 < 0. \label{eq:ds}
\eeq

We first consider the region with  $0<x<R$. From Eq.(\ref{eq:tx}): 
\beq
T'(x)=-\frac{-2xM_g}{(1+x^2)^2}=\frac{2xM_g}{(1+x^2)^2}>0,\label{eq:dt}
\eeq
we have $k'(x)=-S'(x)+T'(x)>0$, and thus $k(x)$
is a monotonic function in this region. 
Due to the singularity at 0 for $T(x)$ and at $R$ for S(x),
we consider the limits and find 
$\lim_{x\to 0^{+}}k(x)=-S(0)+\lim_{x\to 0^{+}}T(x)=-M_g$ and 
$\lim_{x\to R^{-}}k(x)=-\lim_{x\to R^{-}}S(x)+T(R)=\infty$. 
Thus, there is
a unique $x_3\in (0,R)$ such that $k(x_3)=0$. 

Similarly, for the region with $-R<x<0$,
from Eq.(\ref{eq:tx}):
\beq
T'(x)=\frac{-2xM_g}{(1+x^2)^2}>0,\label{eq:dt}
\eeq
so we have $k'(x)=-S'(x)+T'(x)>0$ and $k(x)$
is a monotonic function in this region. 
Because $\lim_{x\to -R^{+}}k(x)=-\infty$  and
$\lim_{x\to 0^{-}}k(x)=M_g>0$,
there is a unique $x_4\in (-R,0)$ such that $k(x_4)=0$.$\Box$

\section{The Bifurcations and Zero-Velocity Curves}

In order to demonstrate the 
analytic results proved above,
we numerically determine the locations of equilibrium points 
by solving $k(x)=0$. 
The locations of the equilibrium points on the x-axis with $m=1$ 
as a function of $M_g$ are shown in Figs.1-2. 
Fig. 1(a) is for $R=0.25$, Fig. 1(b) is for $R=0.5$,
Fig. 2(a) is for $R=1$ and Fig. 2(b) is for $R=2$.

It is clear that when $M_g=0$, there are three equilibrium points, 
L1, L2 and L3 on $x-$axis; when $M_g>0$, 
there are four equilibrium points, 
L2, L3, JY1 and JY2 on $x-$axis.
The separations between these equilibrium points are larger for larger $R$.
Moreover, it is interesting that when $M_g$ increases from 0 to 20,
the separation between L2 and L3 decreases, but 
the separation between JY1 and JY2 increases.

In Fig. 3, the zero-velocity curves, which are obtained through 
Eq.(\ref{eq:Jacobi}), of the case with $m=1$ and $R=1$ 
are presented.  
Fig. 3(a) is for $M_g=1$, Fig. 3(b) is for $M_g=10$,
Fig. 3(c) is for $M_g=30$ and Fig. 3(d) is for $M_g=100$.
The $+$ signs are the locations for Lagrange Points, 
and the squares indicate the Jiang-Yeh Points. 
These points are 
numerically determined by solving $k(x)=0$ and $h(y)=0$.
It is shown that they are completely consistent with the 
locations of equilibrium points implied by zero-velocity curves.  
The locations and the values of Jacobi integral $C_J$ of 
all equilibrium points shown in Fig. 3 are summarized in Table 1.

{\centerline {\bf Table 1. The Locations and $C_J$ of 
Equilibrium Points}}
    \begin{center}
       \begin{tabular}{|c|c|c|c|c|c|c|c|}\hline
\multirow{3}{*}{$M_g=1$}&   & L2 & L3 & L4& L5 & JY1  &JY2 \\ \cline{2-8}
&$(x_e,y_e)$  &(1.94,0)&(-1.94,0) &(0,1.29) &(0,-1.29) & (0.19,0)&
(-0.19,0)\\ \cline{2-8}
&$C_J$  &\multicolumn{2}{|c|}{6.58} &\multicolumn{2}{|c|}{5.02}
&\multicolumn{2}{|c|}{6.94} \\ \hline 
\multirow{3}{*}{$M_g=10$}&  & L2 & L3 & L4& L5 & JY1  &JY2 \\ \cline{2-8}
&$(x_e,y_e)$  &(1.47,0)&(-1.47,0) &(0,1.04) &(0,-1.04) & (0.56,0)&
(-0.56,0)\\ \cline{2-8}
&$C_J$ &\multicolumn{2}{|c|}{28.36} &\multicolumn{2}{|c|}{23.77}
&\multicolumn{2}{|c|}{28.67} \\ \hline 
\multirow{3}{*}{$M_g=30$} && L2& L3 & L4 & L5 & JY1  &JY2 \\ \cline{2-8}
&$(x_e,y_e)$  &(1.33,0)&(-1.33,0) &(0,1.01) &(0,-1.01) & (0.69,0)&
(-0.69,0)\\ \cline{2-8}
& $C_J$ &\multicolumn{2}{|c|}{72.58} &\multicolumn{2}{|c|}{65.20}
&\multicolumn{2}{|c|}{72.90} \\ \hline  
\multirow{3}{*}{$M_g=100$}& & L2 & L3 & L4 &  L5 & JY1  &JY2 \\ \cline{2-8}
&$(x_e,y_e)$  &(1.22,0)&(-1.22,0) &(0,1.0) &(0,-1.0) & (0.79,0)&
(-0.79,0)\\ \cline{2-8}
&$C_J$  &\multicolumn{2}{|c|}{222.11} &\multicolumn{2}{|c|}{210.16}
&\multicolumn{2}{|c|}{222.43} \\ \hline 
\end{tabular}
    \end{center}
\normalsize

\section{The Stability of Equilibrium Points}
After the existence of equilibrium points is confirmed, and 
the locations of equilibrium points are determined, it would be 
interesting to understand the stability around these points.
We now consider the following system: 
\beq \left\{
\begin{array}{ll}
&  \frac{dx}{dt}=u,  \\
& \frac{dy}{dt}=v, \\
& \frac{du}{dt}=2nv +A(x,y),\\
& \frac{d v}{dt}=-2nu+B(x,y),
\end{array}\right. 
\eeq
where $A(x,y)$, $B(x,y)$ are defined in Eqs.(\ref{eq:gf2})-(\ref{eq:gg2}).

To study the stability of equilibrium points, we need to know the
properties of the eigenvalues of equilibrium points.
The characteristic 
equation of the eigenvalue $\lambda$ is  
\beq
\lambda^4+(4n^2-A_x-B_y)\lambda^2+2n(A_y-B_x)
\lambda+A_xB_y-B_xA_y=0, 
\label{eq:labd1}
\eeq
\no where $A_x\equiv {\partial A(x,y)}/{\partial x}$,
$A_y\equiv {\partial A(x,y)}/{\partial y}$,
$B_x\equiv {\partial B(x,y)}/{\partial x}$, and
$B_y\equiv {\partial B(x,y)}/{\partial y}$.
Thus,
\beqn
&&A_x=n^2-\frac{m}{r_1^3}-
\frac{m}{r_2^3}+\frac{3m(x+R)^2}{r_1^5}+\frac{3m(x-R)^2}{r_2^5}
-\frac{M_g}{r(1+r^2)}+ \frac{M_g x^2(1+3r^2)}{r(r+r^3)^2},\label{eq:A_x} \\
& & A_y=\frac{3m(x+R)y}{r_1^5}+\frac{3m(x-R)y}{r_2^5}+
\frac{M_gxy(1+3r^2)}{r(r+r^3)^2}, \label{eq:A_y} \\
& & B_x=\frac{3my(x+R)}{r_1^5}+\frac{3my(x-R)}{r_2^5}+\frac{M_gxy(1+3r^2)}
{r(r+r^3)^2},\label{eq:B_x}\\
& & B_y=n^2-\frac{m}{r_1^3}-\frac{m}{r_2^3}+\frac{3my^2}{r_1^5}
+\frac{3my^2}{r_2^5}-\frac{M_g}{r(1+r^2)}+\frac{M_gy^2(1+3r^2)}{r(r+r^3)^2}.
\label{eq:B_y}
\eeqn

From Eqs.(\ref{eq:A_y}) and (\ref{eq:B_x}), 
for any $x_e$ we have 
$A_y(x_e,0)=B_x(x_e,0)=0$, 
and for any $y_e$ we have 
$A_y(0,y_e)=B_x(0,y_e)=0$.
In order to do further investigation,
some parameters need to be specified:
we set $m=1$ and $R=1$ for all the results in this section.
Thus, from Eq.(\ref{eq:n2})  
we have:
$$n^2=\frac{1}{4}+\frac{M_g}{2}.$$

At first, we consider the equilibrium point L2 and JY1,
$(x_e,y_e)$, which satisfies $k(x_e)=0$ with $x_e>0$ and $y_e=0$.
Due to $A_y(x_e,0)=0$ and $B_x(x_e,0)=0$, 
Eq.(\ref{eq:labd1}) becomes:
\beq
\lambda^4+(4n^2-A_x-B_y)\lambda^2+A_x B_y=0. 
\label{eq:labd2}
\eeq
For convenience, we define $\Omega=A_xB_y$ and
$\Pi\equiv A_x+B_y-4n^2$.
Therefore, we have roots :
\beq
\lambda^2_{+}=\frac{\Pi+\sqrt{\Pi^2-4\Omega}}{2} \quad {\rm and} \quad 
\lambda^2_{-}=\frac{\Pi-\sqrt{\Pi^2-4\Omega}}{2}.
\label{eq:lamda3}
\eeq
Moreover, $A_x(x_e,0)$ and  $B_y(x_e,0)$ can be 
expressed as: 
\beqn
& & A_x(x_e,0)=n^2+\frac{2}{|x_e+1|^3}+\frac{2}{|x_e-1|^3}
+\frac{2x_eM_g}{(1+x_e^2)^2} >0,
\label{eq:Ax_xe} \\
&& B_y(x_e,0)=\frac{1}{x_e}\left(\frac{1}{|{x_e}+1|^3}
-\frac{1}{|{x_e}-1|^3} 
\right) <0
\label{eq:By_xe}
\eeqn
(see Appendix A for details).

For L2, $x_e>R=1$, from Eqs. (\ref{eq:Ax_xe})-(\ref{eq:By_xe}), 
since $A_x(x_e,0)>0$ and $B_y(x_e,0)<0$, we have 
$\Omega=A_x(x_e,0)B_y(x_e,0)<0$. 
Thus, $\Pi^2-4\Omega>0$, and we have $\lambda^2_{+}>0$
and $\lambda^2_{-}<0$. 
As in Szebehely (1967); 
this indicates that it is an unstable equilibrium point.  

For JY1, it has $0<x_e<1$ and $y_e=0$.   
 From  Eqs. (\ref{eq:Ax_xe})-(\ref{eq:By_xe}),  $A_x(x_e,0)>0$ and 
$B_y(x_e,0)<0$, so $\Omega=A_xB_y<0$. 
Thus, $\Pi^2-4\Omega>0$, we have $\lambda^2_{+}>0$
and $\lambda^2_{-}<0$. 
Therefore, JY1 is also an unstable equilibrium point.

Because our system is symmetric with respect to the 
$y$-axis, the above results are also valid 
for L3 and JY2.
Thus, the equilibrium 
points L3 and JY2 are unstable.

Secondly, we study the equilibrium point L4, which can be written as 
$(0, y_e)$ with $y_e > 0$. 
As mentioned previously,
we have $A_y(0,y_e)=B_x(0,y_e)=0$. Thus, 
Eqs.(\ref{eq:labd2}) and (\ref{eq:lamda3})
are also valid here.
We find that:
\beq
A_x(0,y_e)=\frac{6}{(1+y_e^2)^{5/2}}>0,
\eeq
but cannot determine the sign of $B_y(0, y_e)$ analytically here. 
Thus, for each given $M_g$, 
the location of L4, $(0, y_e)$, and then the 
corresponding value of  $B_y(0,y_e)$, 
$\Pi^2-4\Omega$, and  $\Pi$ 
are determined numerically, as shown in Fig. 4(a)-(c).
From Fig. 4(a), we know $B_y(0,y_e)>0$ and 
so $\Omega= A_x B_y >0$. 
Fig. 4(b) shows the value of $\Pi^2-4\Omega$
as a function of $M_g$. It is clear that $\Pi^2-4\Omega$
is not a monotonic function of $M_g$.
There is a critical value $M_{cr}\sim 4.213$, such that
when  $0< M_g <M_{cr}$, we have $\Pi^2-4\Omega < 0$;
when  $M_g \ge M_{cr}$, we have $\Pi^2-4\Omega \ge 0$.
For the case when $\Pi^2-4\Omega < 0$,
both $\lambda^2_{+}$ and $\lambda^2_{-}$ are complex numbers. 
This leads to both $\lambda_{+}$ and $\lambda_{-}$ 
having a root which contains a positive real part, so that 
L4 is unstable. 
For the case when $\Pi^2-4\Omega \ge 0$, we need to know the value of 
$\Pi$. As shown in Fig. 4(c), 
we find that $\Pi<0$ for the considered value of $M_g$. 
We thus have $\lambda^2_{+}<0$ and $\lambda^2_{-}<0$. 
This leads to  both $\lambda_{+}$ 
and $\lambda_{-}$ being pure imaginary numbers, so that 
L4 is a center.

Because our system is symmetric with respect to the $x$-axis,
the above results are also valid for L5.
Thus, the equilibrium 
point L5 is either an unstable point 
or a center.

\section{Concluding Remarks}

We have studied the existence and uniqueness of equilibrium points,
including Lagrange Points and Jiang-Yeh Points, 
of a galactic system with supermassive binary black holes embedded
in a central cuspy galactic halo. 
Due to the cuspy density profile and focusing on 
the case with an equal mass binary black hole, we found 
that the central origin is a singular point.
We also found that the Lagrange Points, L2, L3, L4 and L5, and 
Jiang-Yeh Points, JY1 and JY2, always exist, i.e. there are six equilibrium 
points in the considered system. This differs
from the previous results of non-cuspy galactic 
profiles that Jiang-Yeh Points
only exist under a particular condition (Jiang \& Yeh 2014). 

The stability analysis was performed for these 
equilibrium points.  It is found that the
equilibrium points L2, L3, JY1 and JY2 are
unstable. The equilibrium point L4 (L5) is unstable
when the galactic total mass $M_g < M_{cr}$, and 
is a neutrally stable center when  $M_g \ge M_{cr}$. 
This critical mass $M_{cr}$, which is about 4.213, is therefore 
an important condition for the stability of L4 and L5.
These new results will be employed to investigate the
cores of early-type galaxies in the near future.
 
\section*{Acknowledgment}
We thank the referee for very helpful suggestions.
We are grateful to the National Center for High-performance Computing
for computer time and facilities. 
This work is supported in part 
by the National Science Council, Taiwan, under
Ing-Guey Jiang's
Grants NSC 100-2112-M-007-003-MY3
and Li-Chin Yeh's
Grants NSC 100-2115-M-134-004.

\clearpage
\section*{Appendix A}

The details of the  
calculations of $A_x(x_e,0)$ and $B_y(x_e,0)$ in Section 5 
are presented here. 
If $m=R=1$, then $r_1^2=(x+1)^2+y^2$ and $r_2^2=(x-1)^2+y^2$.
From Eq. (\ref{eq:A_x}),
\beqn
& & A_x(x_e,0)=n^2-\frac{1}{|x_e+1|^3}-\frac{1}{|x_e-1|^3} +\frac{3(x_e+1)^2}
{|x_e+1|^5}-\frac{3(x_e-1)^2}{|x_e-1|^5} \non \\
&& \qquad \qquad \qquad   -\frac{M_g}{|x_e|(1+x_e^2)}
+\frac{M_gx_e^2(1+3x_e^2)}{|x_e|^3(1+x_e^2)^2} \non \\
&&= n^2+\frac{2}{|x_e+1|^3}+\frac{2}{|x_e-1|^3}
-\frac{M_g(1+x_e^2)}{|x_e|(1+x_e^2)^2}+\frac{M_g(1+3x_e^2)}{|x_e|(1+x_e^2)^2}  \non \\
&&= n^2+\frac{2}{|x_e+1|^3}+\frac{2}{|x_e-1|^3}
+\frac{2|x_e|M_g}{(1+x_e^2)^2} >0. \non
\eeqn

On the other hand,
from {\bf Remark B}, since $(x_e,0)$ is an equilibrium point,
$k(x_e)=0$.
By Eq.(\ref{eq:kxe}) and $m=R=1$, we have
\beq
-\frac{M_g}{|x_e|(1+x_e^2)}=\frac{(x_e+1)}{x_e|x_e+1|^3}
+\frac{(x_e-1)}{x_e|x_e-1|^3}-n^2.  \label{eq:kby}
\eeq
 From Eq. (\ref{eq:B_y}) and Eq. (\ref{eq:kby}), we have:
\beqn
&& B_y(x_e,0)=n^2-\frac{1}{|x_e+1|^3}-\frac{1}{|x_e-1|^3}-\frac{M_g}
{|x_e|(1+x_e^2)} \non \\
& &= n^2-\frac{x_e}{x_e|x_e+1|^3}-\frac{x_e}{x_e|x_e-1|^3}
+\frac{(x_e+1)}{x_e|x_e+1|^3}+\frac{(x_e-1)}{x_e|x_e-1|^3}-n^2\non \\
& &=\frac{1}{x_e}\left(\frac{1}{|{x_e}+1|^3}
-\frac{1}{|{x_e}-1|^3}
\right) <0. \non
\eeqn

\clearpage
\begin{figure}[ht]
\centering
\includegraphics[width=0.75\textwidth]{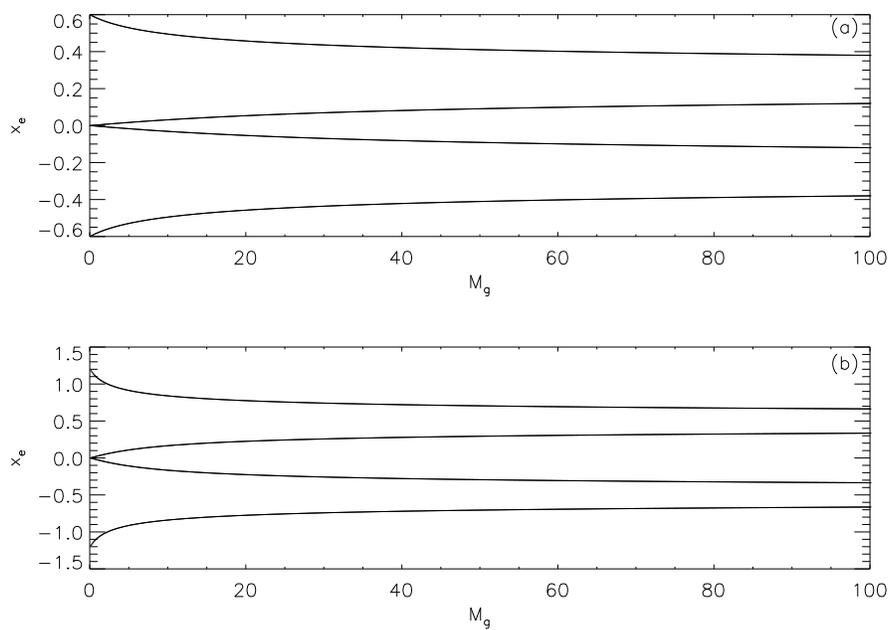}
\caption{The locations of the equilibrium points 
on the $x$-axis with $m=1$
as a function of $M_g$. 
(a) is for $R=0.25$ and (b) is for $R=0.5$.}
\label{fig:fig1}
\end{figure}

\clearpage
\begin{figure}[ht]
\centering
\includegraphics[width=0.75\textwidth]{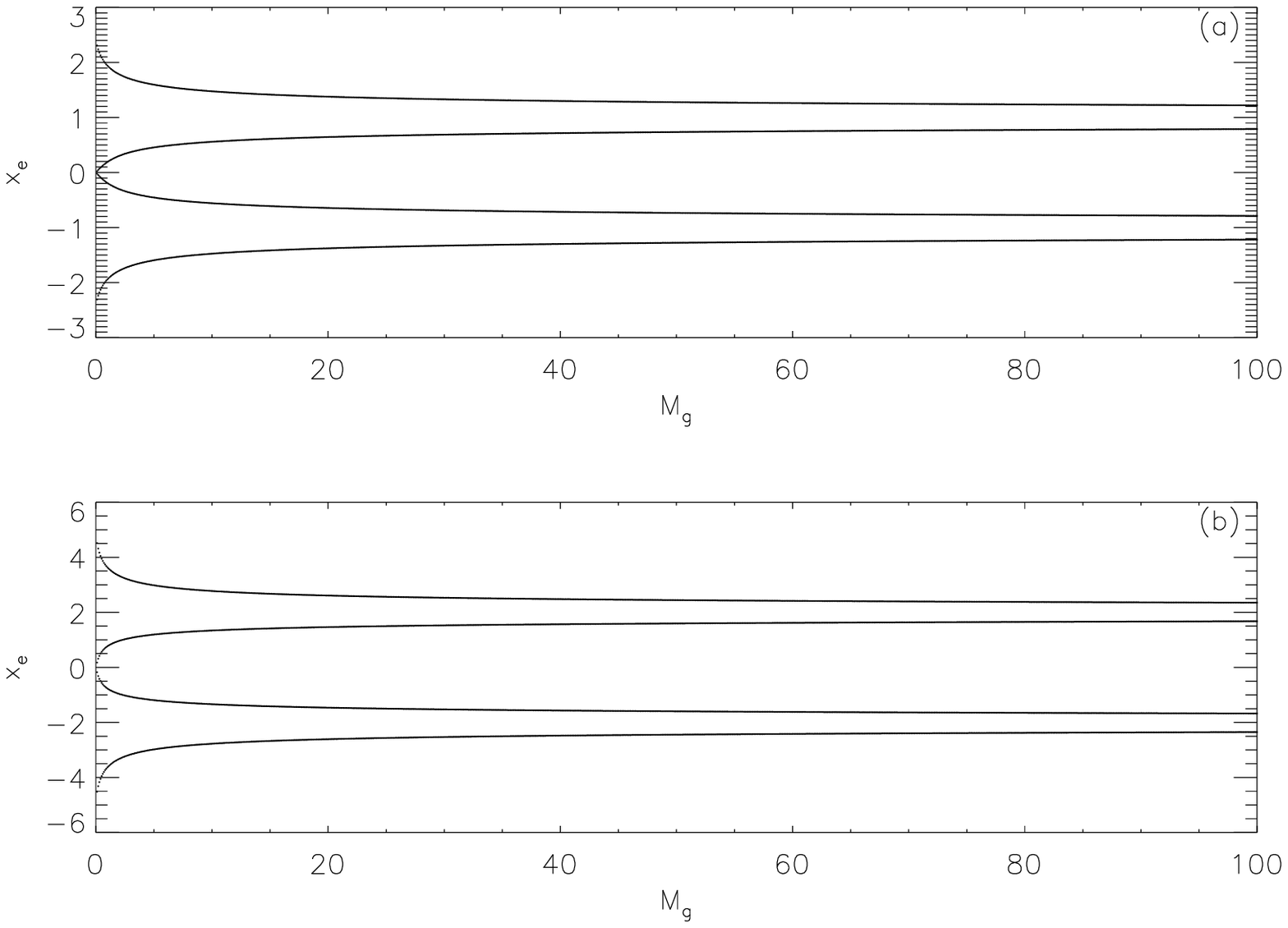}
\caption{The locations of the equilibrium points on the $x$-axis with $m=1$
as a function of $M_g$. (a) is for $R=1$ and (b) is for $R=2$.}
 \label{fig:fig2}
\end{figure}
\clearpage
\begin{figure}[ht]
\centering
\includegraphics[width=1.\textwidth]{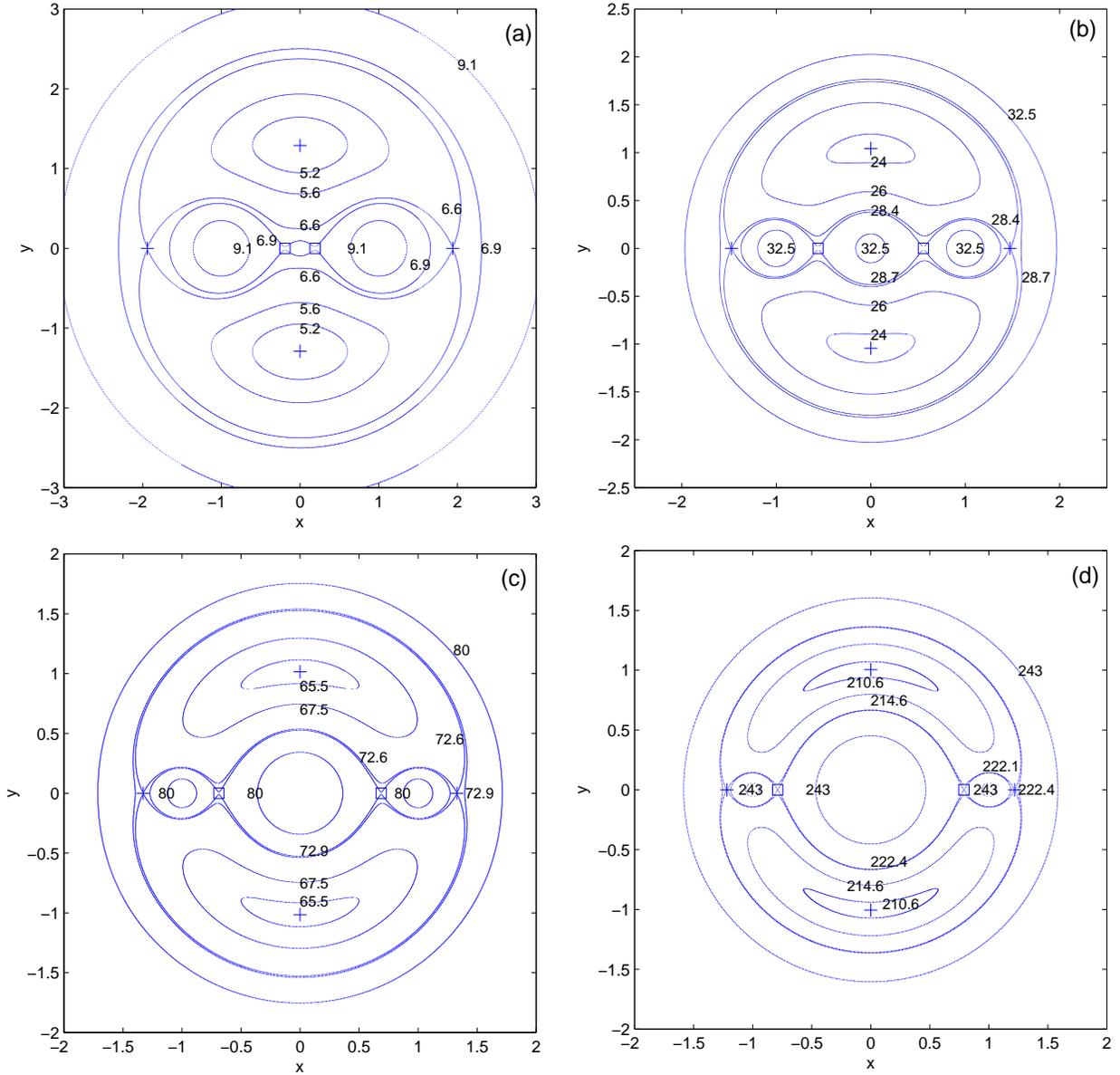}
\caption{The zero-velocity curves of the system when $R=1$ and $m=1$,
on which
the corresponding values of the Jacobi integral $C_J$ are labeled.
(a) is for $M_g=1$, (b) is for $M_g=10$,
(c) is for $M_g=30$, (d) is for $M_g=100$.
The $+$ signs indicate the locations of Lagrange Points
and the squares indicate the locations of Jiang-Yeh Points.
} 
\label{fig:fig3}
\end{figure}

\clearpage
\begin{figure}[ht]
\centering
\includegraphics[width=1.\textwidth]{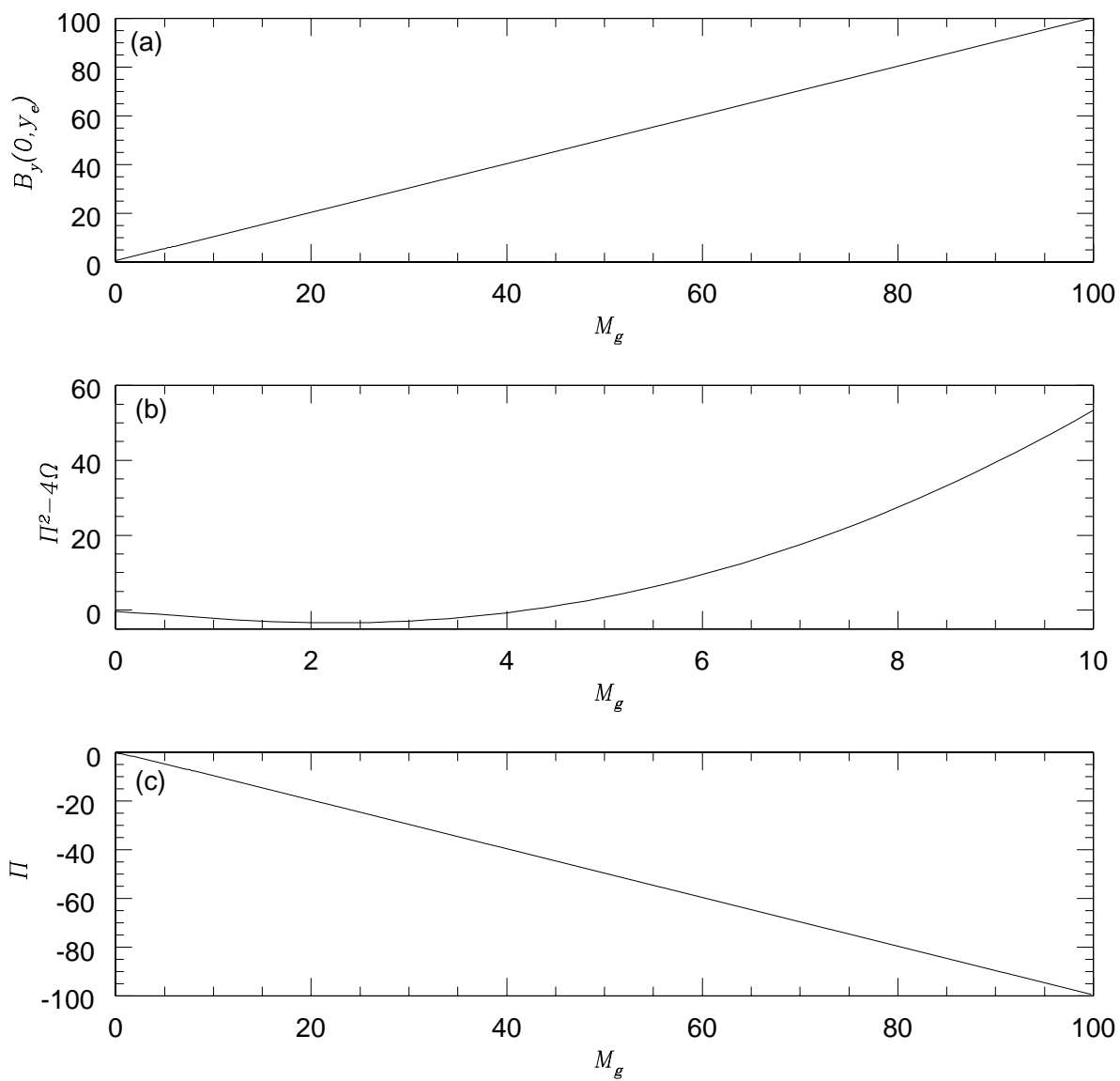}
\caption{
(a) The value of $B_y(0, y_e)$ as a function of $M_g$ for L4.
(b) $\Pi^2 - 4\Omega$ as a function of $M_g$ for L4. 
(c) $\Pi$ as a function of $M_g$ for L4. }
 \label{fig:fig4}
\end{figure}

\end{document}